\documentclass[journal]{IEEEtran}

\usepackage[T1]{fontenc}
\usepackage[utf8]{inputenc}
\usepackage{amsmath,amssymb,amsfonts}
\usepackage{graphicx}
\usepackage{booktabs}
\usepackage{multirow}
\usepackage{array}
\usepackage{cite}
\usepackage{url}
\usepackage{pifont}   
\usepackage{orcidlink}
\usepackage{balance}

\graphicspath{{figures/}}

\newcommand{\Dsem}{D_{\mathrm{sem}}}
\newcommand{\erank}{\operatorname{erank}}
\newcommand{\real}[1]{\mathbb{R}^{#1}}

\begin{document}

\title{AeroLat: Channel-Aware Latent Space Semantic Communication for Decentralized UAV Swarms}

\author{

Rajdeep Ghosh\textsuperscript{\orcidlink{0009-0006-3085-5546}},
\IEEEmembership{Graduate Student Member, IEEE},
Goparaju Venkata Seshachala Sree Vatsava\textsuperscript{\orcidlink{0009-0002-9345-5436}},\\
Sudip Misra\textsuperscript{\orcidlink{0000-0002-2467-6414}},
\IEEEmembership{Fellow, IEEE, ACM}
\thanks{R. Ghosh, G. V. S. S. Vatsava, and S. Misra  are with the Department of Computer Science and Engineering, Indian Institute of Technology Kharagpur,
Kharagpur 721302, India (e-mail: rajghosh2507.24@kgpian.iitkgp.ac.in; goparajusreevatsava@gmail.com; sudipm@iitkgp.ac.in).}%
}

\maketitle
\begin{abstract}
Communication in latent space offers an intriguing alternative to symbolic messages for decentralized autonomous Unmanned Aerial Vehicle (UAV) swarms operating over bandwidth-constrained, time-varying wireless links. However, when homogeneous frozen models are prompted with discretized perceptual inputs, their broadcast states collapse toward the shared prompt template.  In view of this, we propose \textit{AeroLat}, a channel-aware latent semantic communication framework that uses evidence injection. The resulting latent states are then passed through an explicit communication model that encompasses bandwidth-limited serialization, additive noise and information staleness, which facilitates a joint assessment of communication fidelity and swarm-level coordination. Across multi-seed simulations, AeroLat provably remains resilient to codec choice, faults and increasing swarm size. It consistently reproduces the latent-swarm anomaly, while no-whitening controls recover the collapse. In particular, AeroLat is capable of reducing false similarity by 97.5\%.
\end{abstract}

\begin{IEEEkeywords}
Semantic communication, latent-space communication, multi-agent systems, LLMs, decentralized UAV swarms, representational collapse.
\end{IEEEkeywords}

\IEEEpeerreviewmaketitle

\section{Introduction}
\IEEEPARstart{U}{AV} swarms for mission-oriented applications such as post-disaster surveillance and solid waste disposal violations, carry zero-shot perception and lightweight onboard Large Language Model (LLM) reasoners. The open coordination question is how such agents should share their beliefs, thoughts, or perceptions. The pre-dominant answer to it is based on natural language or structured symbolic messaging. However, this imposes what prior work calls a \emph{symbolic bottleneck} --- auto-regressive decoding latency at the sender, quantization of a rich internal belief distribution into one discrete report, and re-encoding ambiguity at the receiver. In coherence with this, recent advancements in this direction of research moved inter-agent communication into the models' continuous latent space. For UAV swarms, the appeal of latent communication extends beyond avoiding the overhead of language generation. When a language model reads text, it builds a vector, a long list of numbers, that represents its internal state: what it currently \emph{has in mind} before it picks a word. Writing a sentence throws most of that away, because the model holds a whole distribution over possible meanings but has to emit one specific string of words. Latent-space communication skips this intermediate symbolic representation by directly exchanging internal model states between agents that share a compatible representation space.

\subsection{Motivation}
Let us consider a mission in which a small fleet surveys an unstructured river basin or urban fringe for waste accumulation and debris. No agent sees the whole scene; each holds a partial, \emph{genuinely ambiguous} view. The fleet must converge on whether an anomaly exists while flying, over a shared wireless medium, on companion computers that also run perception. Three properties of this setting drive every subsequent design decision. First, the \emph{symbolic bottleneck} presents a concrete system-level challenge, due to its measurable overhead in both communication and decision-making. If agents must serialize belief into language before transmitting, three losses compound.In measurements, it takes $0.88$ s per message on a data-center GPU and $ 3.1$ s on an edge-grade GPU, which is much higher than a single forward pass. On the other hand, the latent channel introduces a substantial communication overhead that remains largely unaccounted for.  

\subsection{Contributions}
We propose AeroLat, a latent semantic-communication stack for decentralized UAV swarms. The proposed system treats the broadcast hidden state as a signal to be engineered rather than as a tensor to be passed. Each agent firstly articulates its full zero-shot visual posterior into a frozen onboard LLM, secondly takes the last-layer hidden states as its message, and lastly removes the shared prompt-template subspace with a linear operator that every agent derives independently and identically, training-free, coordination-free, and with one projection per message. The whitened state then crosses an explicitly modeled aerial link with erasures, jitter, bandwidth-limited serialization, noise and quantization, and the task information it carries is measured rather than asserted. Consequently, consensus tracks the scene rather than the prompt, both in emulation and across twenty-five cold-boot flights of the full stack. In brief, the contributions of the proposed work are as follows:
\begin{itemize}
\item We propose \emph{evidence injection} and \emph{template whitening}, a shared linear operator that requires neither training nor inter-agent coordination, but restores discriminative similarity to $0.025$-- a unique contribution to the state-of-the-art in decentralized UAV swarm communication.

\item We benchmark the proposed model against existing dispatching methodologies for swarms, implemented on an identical perception stack, across twenty seeded runs using Welch and Mann--Whitney tests with Holm correction.

\item We validate the complete stack in a PX4/AirSim campaign on an edge-class GPU, spanning codec comparison, fault injection, impaired channels, and fleet scaling to show the efficacy of the proposed work.
\end{itemize}

\begin{table*}[!t]
\caption{Related works across the two capability clusters this problem requires. \checkmark{} = addressed; \ding{55} = not addressed; \emph{p} = partially addressed. The upper block represents latent-space multi-agent communication; the middle block represents semantic communication; the lower block represents UAV networking and mobile computing.}
\label{tab:litreview}
\centering
\footnotesize
\setlength{\tabcolsep}{3.2pt}
\begin{tabular}{llccccccc}
\toprule
& & \multicolumn{3}{c}{\emph{latent-space semantics}} & \multicolumn{4}{c}{\emph{communication \& platform realism}} \\
\cmidrule(lr){3-5}\cmidrule(lr){6-9}
Work & Domain & LLM latent & Anti- & Measured & Channel & Mobility & Physical & Seeded \\
     &        & payload    & collapse & info.\ bounds & model & / node speed & platform & stats \\
\midrule
Interlat~\cite{interlat}          & latent MAS & \checkmark & \ding{55} & \ding{55} & \ding{55} & \ding{55} & \ding{55} & \emph{p} \\
LatentMAS~\cite{latentmas}        & latent MAS & \checkmark & \ding{55} & \emph{p} & \ding{55} & \ding{55} & \ding{55} & \checkmark \\
CIPHER~\cite{cipher}              & latent MAS & \checkmark & \ding{55} & \ding{55} & \ding{55} & \ding{55} & \ding{55} & \emph{p} \\
ThoughtComm~\cite{thoughtcomm}    & latent MAS & \checkmark & \emph{p} & \emph{p} & \ding{55} & \ding{55} & \ding{55} & \checkmark \\
KVComm~\cite{kvcomm}              & latent MAS & \checkmark & \ding{55} & \ding{55} & \ding{55} & \ding{55} & \ding{55} & \checkmark \\
Collapse study~\cite{collapse}    & LLM committees & \checkmark & \emph{p} & \ding{55} & \ding{55} & \ding{55} & \ding{55} & \checkmark \\
\midrule
DeepSC~\cite{deepsc}              & semantic comm. & \ding{55} & \ding{55} & \emph{p} & \checkmark & \ding{55} & \ding{55} & \checkmark \\
Semantics for 6G~\cite{sana6g}    & semantic comm. & \ding{55} & \ding{55} & \emph{p} & \checkmark & \ding{55} & \ding{55} & \emph{p} \\
KG semantic comm.~\cite{kgsemcom} & semantic comm. & \ding{55} & \ding{55} & \ding{55} & \checkmark & \ding{55} & \ding{55} & \checkmark \\
\midrule
Drone squads~\cite{tmc1}          & UAV fleets & \ding{55} & \ding{55} & \ding{55} & \emph{p} & \checkmark & \checkmark & \checkmark \\
CLEVER~\cite{tmc6}                & HetNet video & \ding{55} & \ding{55} & \ding{55} & \checkmark & \checkmark & \ding{55} & \checkmark \\
Cooperative DNN inf.~\cite{tmc7}  & edge inference & \ding{55} & \ding{55} & \ding{55} & \emph{p} & \ding{55} & \emph{p} & \checkmark \\
\midrule
\textbf{AeroLat (proposed work)} & \textbf{UAV latent comm.} & \checkmark & \checkmark & \checkmark & \checkmark & \checkmark & \checkmark & \checkmark \\
\bottomrule
\end{tabular}
\end{table*}

\section{Related Work}\label{sec:relatedwork}

Researchers have been developing new methodologies for a semantic communication stack for decentralized UAV swarms. Table~\ref{tab:litreview} summarizes the existing landscape. In this section, we briefly present some of the existing literature categorized as: 1) Latent Inter-Agent Communication and 2) Semantic Communication and UAV Networking.

\subsection{Latent Inter-Agent Communication}
Interlat~\cite{interlat} established end-to-end communication via last hidden states, with learned compression into short latent prefixes. LatentMAS~\cite{latentmas} demonstrated training-free latent collaboration via last-layer embeddings and shared KV working memory, reporting up to $14.6\%$ accuracy gains and $4$-$4.3\times$ faster inference over text-based multi-agent systems. CIPHER~\cite{cipher} bypasses token sampling with expectation embeddings. ThoughtComm~\cite{thoughtcomm} establishes identifiability guarantees through sparsity-regularized autoencoders, while KVComm~\cite{kvcomm} enables latent communication through the transmission of calibrated KV caches. Dense inter-model messaging without a shared latent geometry is explored in~\cite{densecomm}. Cross-model latent alignment is enabled by relative representations~\cite{relrep} and semantic-alignment translation~\cite{semalign}; continuous-latent reasoning within a single model traces back to COCONUT~\cite{coconut} and CODI~\cite{codi}, and has been extended to vision-language models in~\cite{mcot}. Homogeneous LLM committees with redundant, near-identical states were formalized as a pathology in~\cite{collapse}, along with diversity metrics from the effective-rank family.

\subsection{Semantic Communication and UAV Networking}
DeepSC and its successors~\cite{deepsc} transmit task-relevant semantics with learned joint source-channel coding; \cite{sana6g} frames semantics as a 6G opportunity; \cite{kgsemcom} couples semantic coding with knowledge graphs. Cooperative aerial fleets have been studied for mission class-timely inspection of targets by drone squads under energy and trip constraints, validated in real-field experiments as well as simulation~\cite{tmc1}; that line optimizes \emph{where drones fly}, whereas we address \emph{what they say to each other} once airborne, so the two are complementary layers of the same system. Cross-layer allocation driven by application-layer state~\cite{tmc6} is precisely the design pattern our entropy-gated codec policy would instantiate for semantic traffic. The work \cite{maity2020core} addresses load management in communication channels. The authors of work \cite{11643249} explored the semantics of communication under index modulation via a semantic-aware stream splitting scheme. In the work \cite{abdelhady2025optimization}, the authors explored a Hybrid architecture for UAVs powered either solely by a laser or solely by a battery to improve reach and durability. Finally, cooperative DNN inference through device placement and model partitioning across heterogeneous edge nodes~\cite{tmc7} addresses the bottleneck our embodied scaling campaign independently identifies. The work in \cite{bithas2020uav} presents a new channel model that simultaneously accounts for the effects of mobility and shadowing. 

\textit{Synthesis}: After a detailed analysis of the existing methodologies, there exists an unresolved gap at the intersection of strands of communicating informative, non-collapsed LLM latent states over constrained mobile channels while sustaining task-relevant semantics in an embodied decentralized swarm during operation. Current approaches to latent multi-agent systems demonstrate that hidden states, expectation embeddings, or KV representations can replace explicit text and reduce inference overhead, but largely assume lossless, co-located software channels and do not examine mobility, bandwidth, packet loss, or information staleness. Studies of networking for UAVs model constraints on mobility, contention, energy, and edge computing, but treat payload semantics as opaque.

\section{System Model}\label{sec:sysmodel}

\begin{table}[!t]
\caption{LIST OF SYMBOLS}
\label{tab:notation}
\centering
\footnotesize
\setlength{\tabcolsep}{4pt}
\begin{tabular}{|l|l|}
\hline
Symbol & Description \\
\hline
$\mathcal{D}_i$ & agent (drone) $i$ \\
$T_{\mathrm{inf}}$ & perception/inference interval ($3$\,s) \\
$I_i(t)$, $v_i(t)$ & frame and CLIP image embedding of $\mathcal{D}_i$ \\
$p_i(t)$ , $H_i(t)$ & CLIP posterior over $C{=}13$ classes; its entropy\\
$x_i(t)$ & evidence-verbalized prompt \\
$h_i(t)$ & raw latent from last $k{=}8$ hidden states in $\real{7168}$ \\
$\mu$ & whitening mean \\
$P$ & top-$m$ template subspace \\
$\psi_i(t)$ & whitened broadcast state \\
$\hat\psi_j$ & peer state as received (post-channel) \\
$\lambda$ & fusion coefficient \\
$a_{jm}$ & attention weight \\
$A_m$ & message age \\
$T_a$ & staleness discount constant \\
$Q_b(\cdot)$ & quantizer at $b$ bits/dim \\
$n_B$ & payload bytes \\
$p_\ell$ & erasure probability \\
$\gamma$ & per-symbol SNR \\
$R_{\mathrm{tot}}$ & shared medium budget \\
$R_{\mathrm{link}}$ & per-link rate \\
$\Lambda(N)$ & aggregate offered mesh load \\
$\Dsem$ & semantic distortion $1-\cos(\psi,\hat\psi)$ \\
$C(t)$ & whitened consensus \\
$v$ & node speed \\
$f_d$ & Doppler shift \\
$T_c$ & coherence time \\
$\tau$ & latency \\
$\rho$ & information density \\
\hline
\end{tabular}
\end{table}

\begin{figure*}[!t]
\centering
\includegraphics[width=0.8\textwidth]{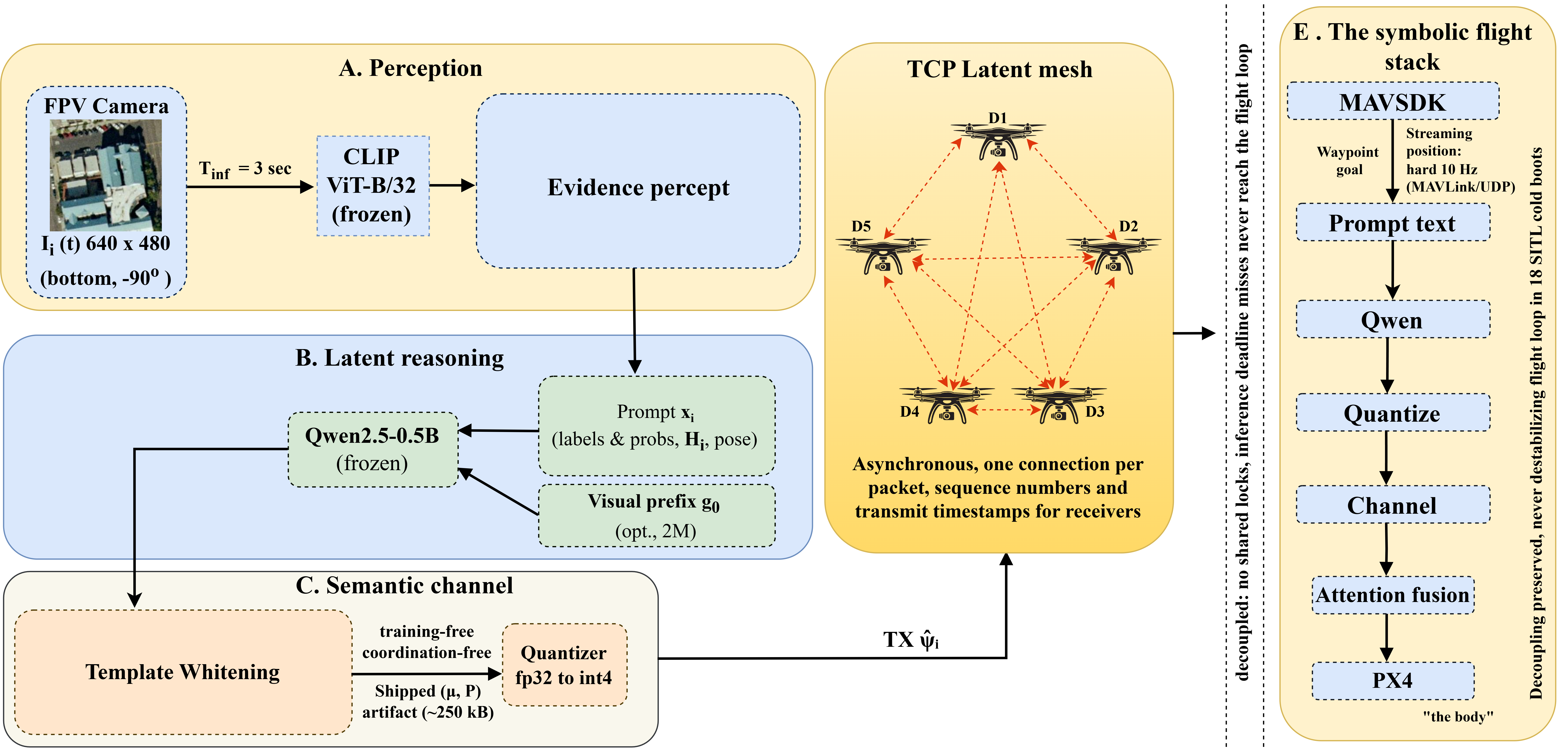}
\caption{AeroLat system architecture for one agent $\mathcal{D}_i$. (A) zero-shot CLIP perception emits the full posterior, image embedding, visual entropy, and pose rather than an $\arg\max$ label; (B) that evidence is verbalized into the prompt of the frozen LLM; (C) template whitening removes the shared common-mode subspace. The centre part depicts the asynchronous TCP latent mesh, one connection per packet with sequence numbers and transmit timestamps so receivers can measure message age. (D) staleness-aware attention fusion; (E) the symbolic flight stack.}
\label{fig:arch}
\end{figure*}

\subsection{Platform and Timing}
Each UAV $i\in\{1,\dots,N\}$, where $N$ is the swarm size, runs a PX4~\cite{px4} flight controller consuming $10$\,Hz position setpoints over MAVLink/UDP, an perception--reasoning stack, CLIP ViT-B/32~\cite{clip} for zero-shot anchoring, and Qwen2.5-0.5B~\cite{qwen} for latent generation on a rolling inference interval $T_{\mathrm{inf}}=3$\,s; and a TCP mesh socket for the semantic side-channel. The two loops share no locks, and an inference deadline miss degrades only message freshness, never flight stability. This decoupling is the architectural invariant of the proposed work as depicted in Fig.~\ref{fig:arch}.

\subsection{Message Generation}
At each inference tick, drone $i$ captures frame $I_i(t)$, computes the CLIP posterior $p_i(t)\in\Delta^{C-1}$ over $C{=}13$ scene/anomaly classes and the image embedding $v_i(t)\in\real{512}$, builds an evidence-verbalized prompt $x_i(t)$, and extracts the last-layer hidden states of the final $k{=}8$ tokens:
\begin{equation}
h_i(t)=\operatorname{vec}\!\big(H^{(L)}_{i,\,-k:}\big)\in\real{7168},\qquad d_{\mathrm{model}}=896 .
\end{equation}
The broadcast state is the whitened residual $\psi_i(t)$. Payload sizes measured on the wire (16-B header included) are $28{,}688$\,B (fp32), $14{,}352$\,B (fp16), $7{,}184$\,B (int8), and $3{,}600$\,B (int4) per state.

\subsection{Channel Model}\label{sec:channelmodel}
Each directed link $(i\!\to\!j)$ applies, in order:
\begin{enumerate}
\item \emph{Source coding}: quantization $Q_b(\cdot)$ at $b\in\{32,16,8,4\}$ bits per dimension (symmetric per-vector scaling for integer modes), with optional top-$s$ magnitude sparsification charged an explicit 32-bit index cost per retained coefficient.

\begin{equation}
R_{\mathrm{link}}=R_{\mathrm{tot}}/\!\left(N(N{-}1)\right)
\end{equation}

\item \emph{Noise}: for analog transmission, Additive White Gaussian Noise (AWGN) at per-symbol SNR $\gamma$, i.e., 
\begin{equation}
\hat\psi=\psi+w, w\sim\mathcal{N}(0,s^2 I)
\end{equation} 
with $s^2=\mathbb{E}[\psi^2]/10^{\gamma/10}$

\item \emph{Staleness}: the receiver discards states older than an age-of-information threshold $A_{\max}$, and fusion discounts survivors by $e^{-A/T_a}$.
\end{enumerate}

The offered load of the full mesh is $\Lambda(N)=N(N{-}1)\,8n_B/T_{\mathrm{inf}}$; with fp32 payloads this evaluates to $0.5$, $1.5$, $6.9$, and $29.1$\,Mb/s at $N{=}3,5,10,20$. Quadratic growth is stated plainly here because it motivates both quantization and learned compression.

\subsection{Complexity and Overhead}\label{sec:complexity}
For each agent and at every inference tick, the computational cost consists of one CLIP forward pass and one LLM forward pass. The whitening projection draws a complexity of $O(mD)$ with $m{=}4$ and $D{=}7168$. The incurred cost is negligible relative to the transformer forward passes and is measured at ${<}1$\,ms. Apart from these, quantization requires $O(D)$ operations and fusion incurs $O(N D)$ complexity, since attention is performed over at most $N{-}1$ peer states. The per-agent computation is therefore $O(ND)$ beyond the two fixed forward passes, and is independent of mission length because no history is retained. Communication cost is one broadcast of $n_B$ bytes per tick per agent, giving the aggregate offered load $\Lambda(N)=N(N{-}1)\,8n_B/T_{\mathrm{inf}}$ as stated above.

\subsection{Metrics}\label{sec:metrics}
The communication quality, task performance, and swarm-state diversity is evaluated using the following metrics.
For \(M\) evaluated transmissions (or mission episodes), the semantic-relay
success rate is
\begin{equation}
    \mathrm{SR}
    =
    \frac{1}{M}
    \sum_{m=1}^{M}
    \mathbf{1}\!\left\{\mathcal{S}_m\right\},
\end{equation}
where \(\mathcal{S}_m\) denotes satisfaction of the task-specific success criterion in episode \(m\). Coordination latency is calculated as,
\begin{equation}
    \tau
    =
    t_{\mathrm{enc}}
    +
    t_{\mathrm{ch}}
    +
    t_{\mathrm{int}},
\end{equation}
where the three terms denote encoding, channel, and receiver-side interpretation time, respectively. We access the efficacy of transmitted bits in preserving the task-relevant information by evaluation the information density as,
\begin{equation}
    \rho
    =
    \frac{\hat I(\hat{\psi};Y)}{b},
\end{equation}
where \(\hat I(\hat{\psi};Y)\) is the estimated mutual information between the received semantic state and task variable \(Y\), and \(b\) is the
number of transmitted bits. We additionally measure the breadth of the receiver's decoded hypothesis.
Let \(q_{(1)} \ge q_{(2)} \ge \cdots \ge q_{(K)}\)
denote the decoded class probabilities in descending order. In order we can define this as,
\begin{equation}
    P_{50}
    =
    \min\left\{
        k :
        \sum_{j=1}^{k} q_{(j)} \ge \frac{1}{2}
    \right\}.
\end{equation}
Thus, smaller \(P_{50}\) indicates a more concentrated decoded belief,
whereas larger \(P_{50}\) indicates greater hypothesis breadth.

Consensus $C(t)$ is the mean off-diagonal entry of the $N\times N$ Gram matrix
$S_{ij}=\cos\!\left(\psi_i(t),\psi_j(t)\right)$ of whitened states. Finally, we
monitor latent-state collapse using effective rank. Let
$\Psi(t)=[\psi_1(t),\dots,\psi_N(t)]^{\!\top}\in\mathbb{R}^{N\times D}$ be the
stacked swarm latent set, $\bar{\Psi}(t)$ its column-centered form, and
$s_1\ge\cdots\ge s_R>0$ the nonzero singular values of $\bar{\Psi}(t)$.
With $\pi_r = s_r / \sum_q s_q$ so that $\pi_r\ge0$ and $\sum_r \pi_r = 1$, the effective rank is
\begin{equation}
    \erank
    =
    \exp\!\left(
        -\sum_r \pi_r \ln \pi_r
    \right)
\end{equation}
 The process costs one degree of freedom, giving $\erank\in[1, N{-}1]$. The measure approaches $1$ when the swarm states collapse to a common direction and approaches $N{-}1$ when they span comparably weighted, independent directions.

\section{The Collapse Pathology and the AeroLat Fix}\label{sec:collapse}

\subsection{Template Whitening}\label{sec:whitening}
Let $\{h_b(t)\}_{b=1}^{B}$ be a calibration buffer of raw latents over diverse scenes. With mean $\mu$ and the top-$m$ right singular vectors $P\in\real{m\times D}$ of the centered buffer, every agent broadcasts
$P$ spans the template subspace, and the residual carries the scene-specific variation. Because all agents run identical frozen weights and a deterministic calibration protocol, $(\mu,P)$ agree across the fleet \emph{without coordination}. All similarity, consensus, fusion, and channel-distortion computations operate on $\psi$. With $m{=}4$, the operator costs one $D\times m$ projection per message.

\subsection{Fusion Under Impairments, and a Second Collapse Mechanism}\label{sec:fusion}
Receiver $j$ fuses its own $\psi_j$ with peer states via scaled dot-product attention \emph{including self}, discounted by message age $A_m$:
\begin{align}
a_{jm} &\propto \exp\!\big(\psi_j^{\top}\psi_m/\sqrt{D}\big)\, e^{-A_m/T_a},\nonumber\\
\psi_j^{\mathrm{new}} &= \mathrm{norm}\Big((1-\lambda)\,\psi_j+\lambda\!\!\sum_{m\neq j}\tilde{a}_{jm}\,\psi_m\Big),
\label{eq:fusion}
\end{align}
The fusion coefficient \(\lambda\) controls the contribution of peer information and is selected empirically. One structural requirement here is that fusion must use the \emph{fresh} percept latent as $\psi_j$, not the previous fused state. Iterating~\eqref{eq:fusion} on fused states is a DeGroot averaging process that converges to a common vector for any $\lambda>0$ regardless of scene diversity. This leads to a dynamical collapse that is distinct from prompt-induced representational collapse. To avoid repeated averaging across rounds, AeroLat re-initializes the local percept every $T_{\mathrm{inf}}$.

\section{Semantic Rate--Distortion and Information Accounting}\label{sec:channel}

Before assessing the channel's performance, we first characterize the impact of transmission impairments on the geometry and task-relevant information carried by the latent state. The following section consequently analyses AeroLat using complementary distortion and information measures against the considered channel operating points.

\subsection{Empirical Rate--Distortion}\label{sec:channelresults}
\begin{figure}[!t]
\centering
\includegraphics[width=0.8\columnwidth]{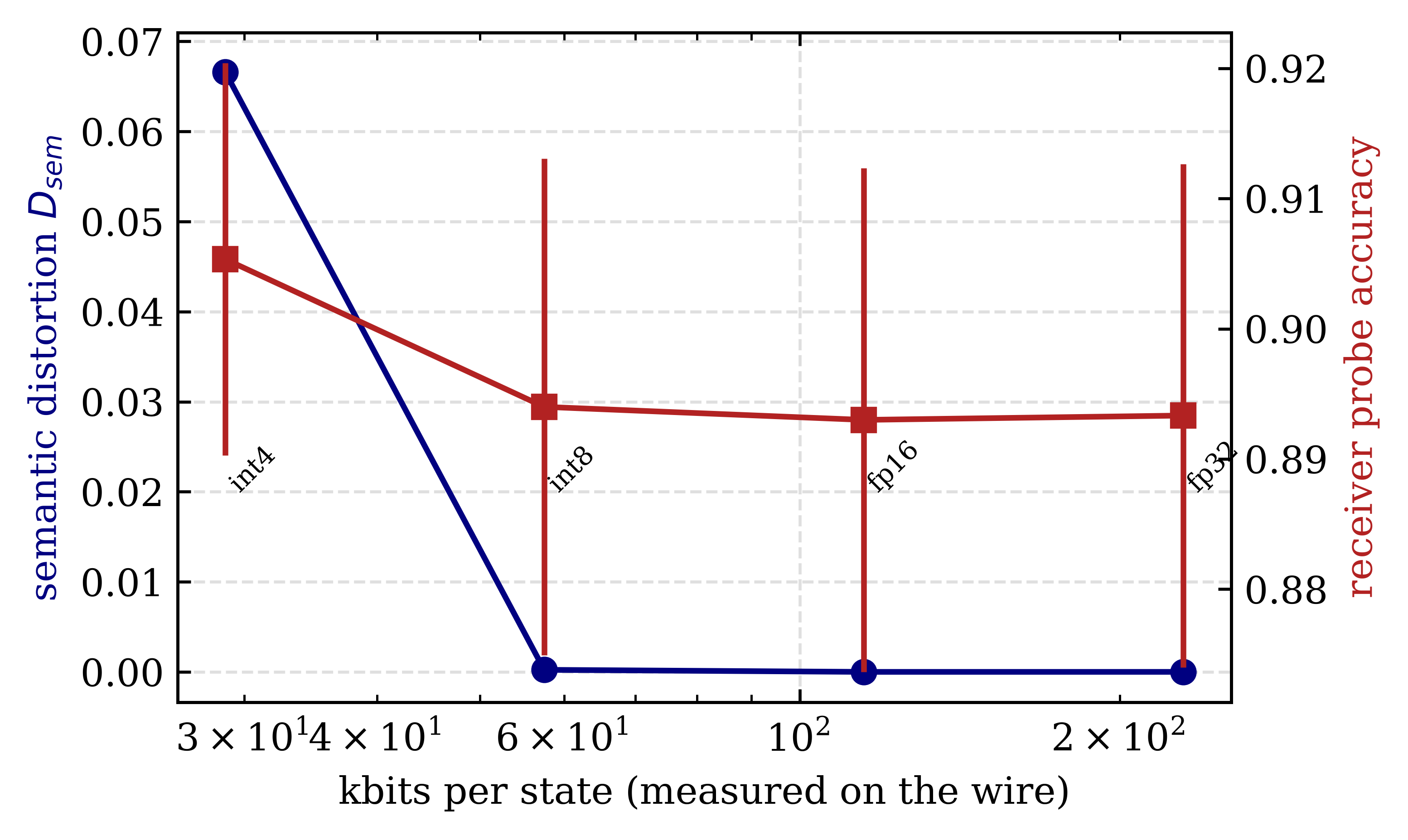}
\caption{Semantic rate--distortion over the real latent corpus (ten seeds). Dense quantization from fp32 ($229.5$\,kb/state) to int8 ($57.5$\,kb) is semantically lossless ($\Dsem\le2\times10^{-4}$); int4 ($28.8$\,kb) costs $\Dsem=0.067$ in geometry while receiver probe accuracy remains $0.905$: within this codec family, rate can be cut $8\times$ before task distortion begins.}
\label{fig:rd}
\end{figure}
\begin{figure}[!t]
\centering
\includegraphics[width=0.8\columnwidth]{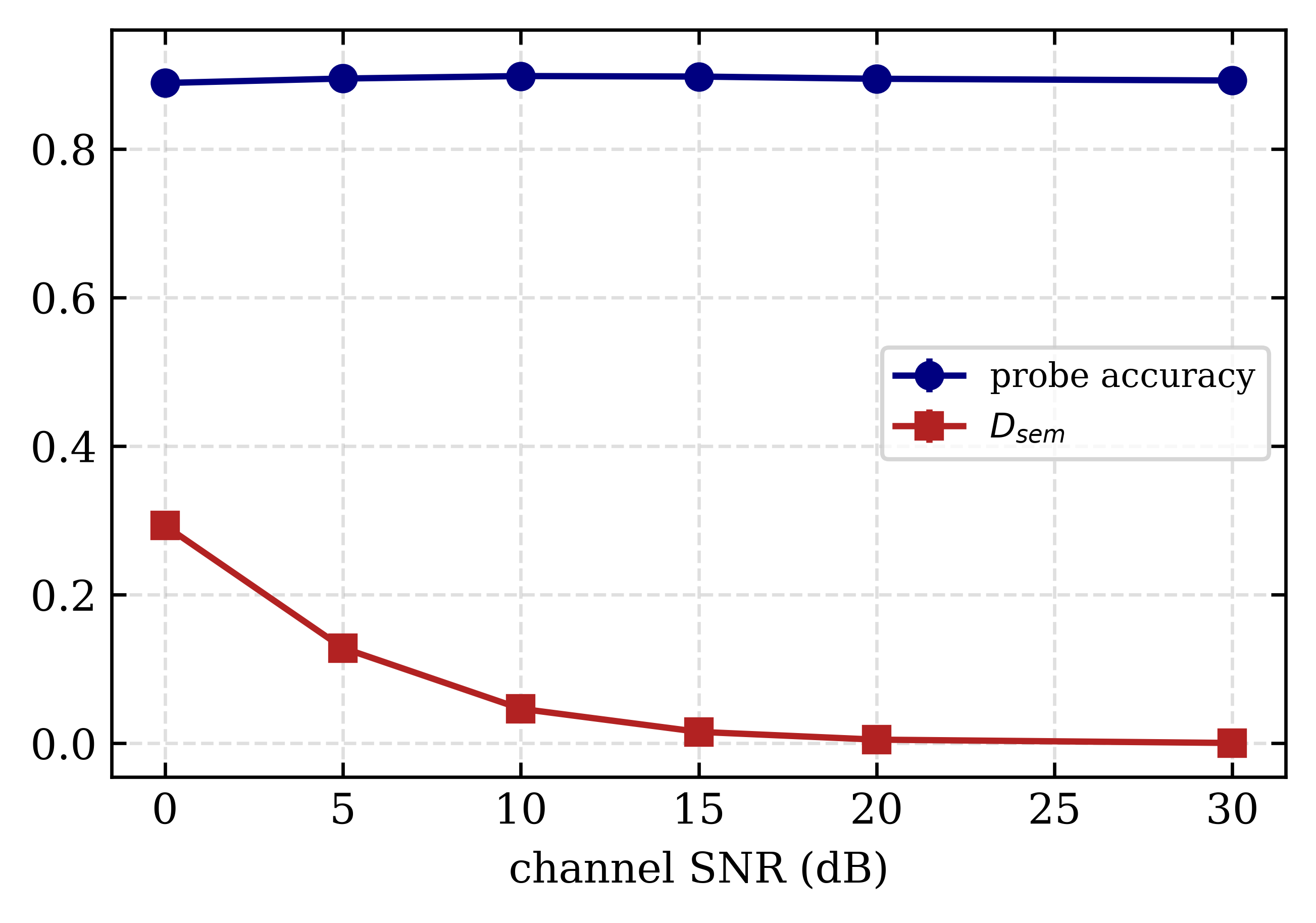}
\caption{AWGN robustness. From 30~dB down to 0~dB SNR, the geometric distortion grows ($\Dsem\,0.000\!\to\!0.293$) while receiver probe accuracy is flat ($0.892\!\to\!0.889$): the whitened latent code degrades gracefully, with no digital cliff.}
\label{fig:awgn}
\end{figure}
Sweeping quantization and sparsity over the corpus yields Fig.~\ref{fig:rd}. Top-$s$ sparsification is reported honestly as \emph{not} competitive at these dimensions once the 32-bit index overhead is included. Generally, for below ${\sim}4\times$ compression, a sparse int8 vector costs more wire bytes than a dense one. AWGN robustness appears in Fig.~\ref{fig:awgn}. The flat accuracy down to 0~dB is the graceful-degradation profile that a digital symbolic packet, catastrophic under residual bit errors, cannot exhibit.

\subsection{Measured Information Density}\label{sec:info}
We quantify the task-relevant information preserved in the received latent state $\hat\psi$ using two lower bounds on $I(\hat\psi;Y)$, with $Y$ the scene/anomaly class: a \emph{Fano} bound from probe error $e$,
\begin{equation}
I \;\ge\; H(Y) - H_b(e) - e\,\log_2\!\big(|\mathcal{Y}|-1\big),
\label{eq:fano}
\end{equation}
and a neural-critic \emph{InfoNCE} bound $I \ge \log N_b - \mathcal{L}_{\mathrm{NCE}}$ trained per operating point. Table~\ref{tab:mi} reports both together with $\rho=\hat{I}/b$. Quantization is nearly free in task information (Table~\ref{tab:mi}), so wire density improves $8\times$ at no measured semantic cost.

\begin{table}[!t]
\caption{Measured information per state and information density $\rho$ (neural InfoNCE and Fano bounds; task ceiling $H(Y)=2.52$ bits; the marginal overshoot at 0~dB reflects finite-batch estimator variance).}
\label{tab:mi}
\centering
\begin{tabular}{|l|c|c|c|c|}
\hline
Operating point & bits/state & $\hat{I}_{\mathrm{NCE}}$ & Fano & $\rho$ (b/kb) \\
\hline
fp32, noiseless & 229{,}504 & $\ge2.52$ & 1.91 & 0.011 \\
fp16            & 114{,}816 & $\ge2.52$ & 1.91 & 0.022 \\
int8            & 57{,}472  & $\ge2.52$ & 1.90 & 0.044 \\
int4            & 28{,}800  & $\ge2.52$ & 1.88 & \textbf{0.088} \\
fp32 @ 10\,dB   & 229{,}504 & $\ge2.52$ & 1.83 & 0.011 \\
int8 @ 10\,dB   & 57{,}472  & $\ge2.52$ & 1.83 & 0.044 \\
fp32 @ 0\,dB    & 229{,}504 & $\ge2.53$ & 1.78 & 0.011 \\
\hline
\end{tabular}
\end{table}

\begin{figure}[!t]
\centering
\includegraphics[width=0.8\columnwidth]{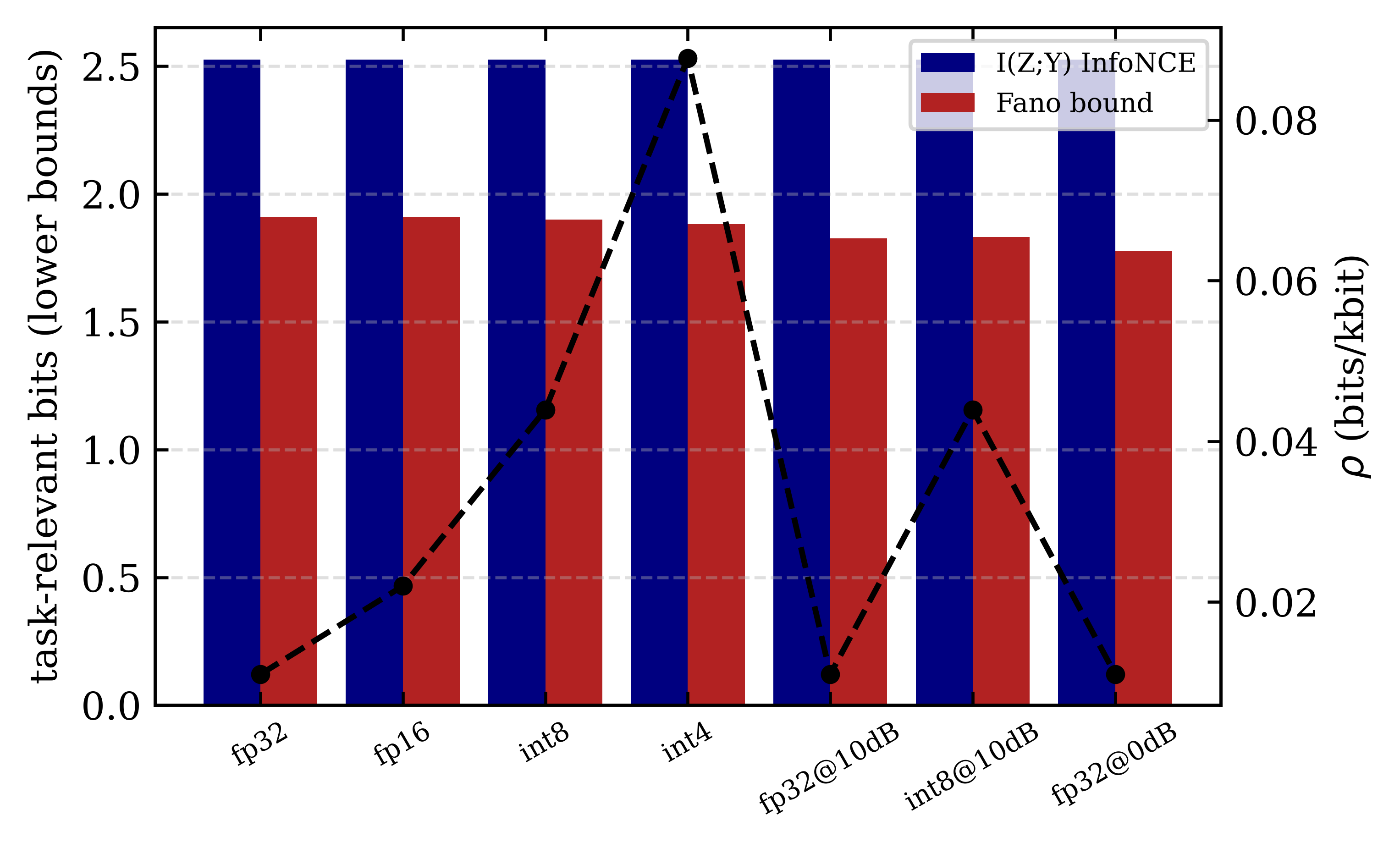}
\caption{Task-relevant information (InfoNCE and Fano lower bounds, bars) and density $\rho$ (line) per channel operating point. The state carries the entire task entropy essentially and retains it under int4 quantization and 0~dB AWGN simultaneously.}
\label{fig:mi}
\end{figure}

\section{Learned Compression: the Four-Loss Protocol}\label{sec:compressor}
The four-loss objective below follows the taxonomy introduced by
Interlat~\cite{interlat}; our contribution here is empirical---prior work states these losses but does not train or dissect them, and the leave-one-loss-out ablations that follow assign each term a measured role. The compressor $M_{\mathrm{c}}$ maps the full latent trajectory $H_L$ to $H_K\in\real{8\times896}$ and is trained with the frozen LLM as measurement device:
\begin{equation}
\mathcal{L}=\mathcal{L}_{\mathrm{task}}+a\,\mathcal{L}_{\mathrm{sep}}+b\,\mathcal{L}_{\mathrm{pref}}+c\,\mathcal{L}_{\mathrm{geom}},
\end{equation}
where $\mathcal{L}_{\mathrm{task}}$ is the cross-entropy of the frozen receiver producing the correct structured survey report conditioned on $H_K$ as a soft prefix. $\mathcal{L}_{\mathrm{sep}}=-\mathrm{JS}\!\left(p_\theta(\cdot\,|\,C,H),\,p_\theta(\cdot\,|\,C,\tilde{H})\right)$ with batch-mismatched $\tilde{H}$ (the ``must-listen'' pressure). $\mathcal{L}_{\mathrm{pref}}$ is a confidence-weighted teacher--student KL between full-trajectory and compressed conditioning. Lastly, $\mathcal{L}_{\mathrm{geom}}=1-\cos(\bar{z}^{(A)},\bar{z}^{(D)})$ aligns step-averaged actor features. As shown in Fig.~\ref{fig:comp} and Table~\ref{tab:comp}, training converges in less than 38 epochs, and the Jensen–-Shannon separation between matched and mismatched prefixes increases by approximately $0.1$ nat, which confirms dependence on the transmitted latent. Ablations show that \(L_{\mathrm{pref}}\) prevents shortcut learning, \(L_{\mathrm{geom}}\) preserves latent orientation, and \(L_{\mathrm{sep}}\) enforces prefix dependence with little effect on task loss. The optional compressor reduces the int8 payload from \(57.5\) to \(14.3\) kb.As shown in Fig.~\ref{fig:comp} Ablations show that \(L_{\mathrm{pref}}\) prevents shortcut learning, \(L_{\mathrm{geom}}\) preserves latent orientation, and \(L_{\mathrm{sep}}\) enforces prefix dependence with little effect on task loss. The optional compressor reduces the int8 payload from \(57.5\) to \(14.3\) kb. The visual-prefix projector further improves the accuracy of validation probes to \(0.973\).

\begin{table}[!t]
\caption{Compressor training and leave-one-loss-out ablations: initial $\to$ final loss values (full run: 30 epochs; ablations: 15; for parity the full run reads $\mathcal{L}_{\mathrm{task}}{=}0.405$ at epoch 14).}
\label{tab:comp}
\centering
\footnotesize
\setlength{\tabcolsep}{3pt}
\begin{tabular}{|l|c|c|c|c|}
\hline
Run & $\mathcal{L}_{\mathrm{task}}$ & $\mathcal{L}_{\mathrm{sep}}$ & $\mathcal{L}_{\mathrm{pref}}$ & $\mathcal{L}_{\mathrm{geom}}$ \\
\hline
full     & $2.37\!\to\!0.39$ & $-.01\!\to\!-.11$ & $3.36\!\to\!1.52$ & $.38\!\to\!.07$ \\
no-sep   & $\to0.39$ & (off) & $\to1.55$ & $\to.08$ \\
no-pref  & $\to\mathbf{0.008}$ & $\to-.13$ & (off) & $\to.13$ \\
no-geom  & $\to0.41$ & $\to-.10$ & $\to\mathbf{1.87}$ & (off) \\
\hline
\end{tabular}
\end{table}

\section{Experimental Evaluation}\label{sec:eval}
This section assesses AeroLat across representation quality, channel robustness, baseline performance, and swarm scalability in both controlled and full-stack settings. For the simulation phase, we utilized a high-fidelity Software-In-The-Loop (SITL) architecture that interfaces with a detailed AirSim (Unreal Engine) neighborhood environment \cite{airsim}, providing a continuous, unstructured 3D testing ground.
\subsection{Protocol}
We construct the evaluation corpus from 1,000 aerial images across six scene classes in the UC Merced Land Use Dataset~\cite{ucmerced}, a benchmark of high-resolution aerial imagery for land-use classification.The fitted whitener is exported as the deployment artifact. We perform semantic-layer experiments (probe evaluation, channel sweeps, ablations, and scaling studies) over 10--20 random seeds and report mean~$\pm$~standard deviation with 95\% bootstrap confidence intervals. Baseline comparisons are done with Welch's $t$-test, Mann–-Whitney $U$-test with Holm correction, and Cohen’s $d$. Full-stack validation is performed using SITL missions on an NVIDIA T1000 (4\,GB), and semantic-layer experiments and training are conducted on an RTX~6000~Ada, on which Airsim runs.

\subsection{Collapse and Representation Diagnostics}\label{sec:beforeafter}
\begin{table}[!t]
\caption{The collapse, reproduced and repaired ($n{=}400$ per condition; chance $=0.167$). Row~4 is the earlier draft's flagship ``$1.00$ similarity,'' explained.}
\label{tab:beforeafter}
\centering
\begin{tabular}{|l|c|c|}
\hline
Metric & Argmax prompt & Evidence prompt \\
 & (prior system) & (AeroLat) \\
\hline
Probe accuracy, raw $h$        & $0.593\pm0.071$ & $\mathbf{0.923\pm0.020}$ \\
Probe accuracy, whitened $\psi$      & $0.500\pm0.065$ & $\mathbf{0.878\pm0.037}$ \\
Fano MI (raw), bits                  & $0.61$          & $\mathbf{1.96}$ \\
Off-diag.\ cosine, raw               & $\mathbf{0.9978}$ & $0.9989$ \\
Off-diag.\ cosine, whitened          & $0.184$         & $\mathbf{0.025}$ \\
Effective rank, raw $\to$ whitened   & $7.9\to4.1$     & $\mathbf{164\to233}$ \\
\hline
\end{tabular}
\end{table}
The collapse and its mitigation are summarized in Table~\ref{tab:beforeafter}. Under argmax prompting, the mean cosine similarity between raw latents across scenes is \(0.9978\). Template whitening is what restores discriminative similarity to $0.025$ across mixed scenes while increasing latent diversity (\(\erank=233\)), indicating that semantic content and representation geometry constitute distinct failure modes. We additionally test the representation on a six-class aerial corpus with \(H(Y)=2.52\) bits and 1{,}000 samples, with three diagnostics:

\begin{itemize}
    \item \textit{D1 (probe recoverability).} Five-fold linear-probing produces $0.915\pm0.032$ accuracy on raw states and $0.916\pm0.030$ accuracy after whitening, in comparison to $0.167$ chance accuracy predicting the scene class from the broadcast vector. 
    \item \textit{D2 (representational sensitivity).} With the template fixed and only vision varied, the ratio of between-scene to within-scene latent distances is $1.34$. 
    \item \textit{D3 (conflict fusion/superposition).} Fusing whitened latents from two \emph{conflicting} scenes at $\lambda{=}0.4$, the fused state's probe top-2 contains \emph{both} source hypotheses in $92.75\%$ of $400$ trials (snapping to a single hypothesis: $7.25\%$).
\end{itemize}

\subsection{Baseline Comparison}\label{sec:baselines}

\begin{figure*}[!t]
\centering
\includegraphics[width=0.8\textwidth]{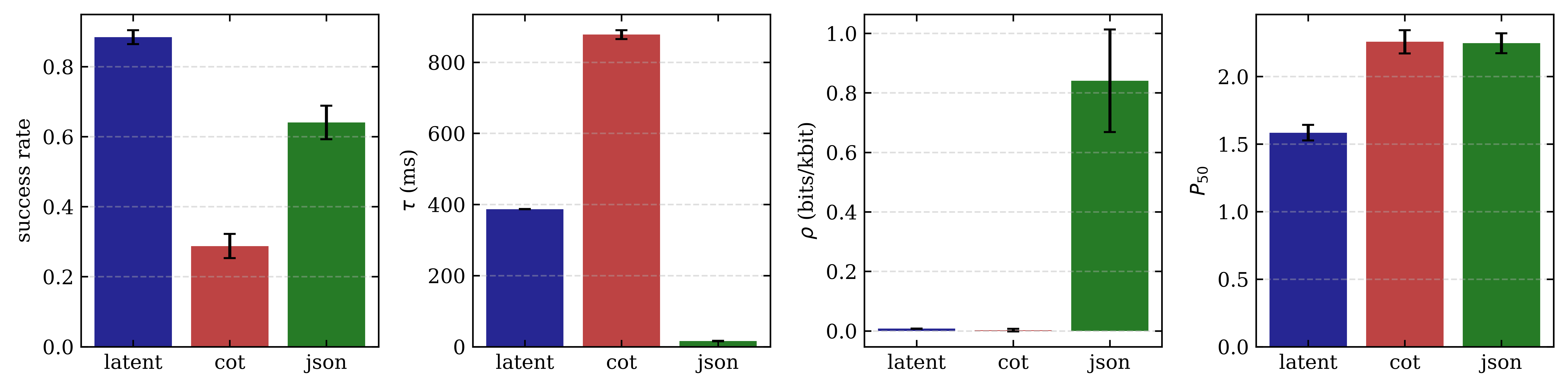}
\caption{Baseline comparison across twenty seeds. Left to right: relay success rate, coordination latency $\tau$, information density $\rho$, and hypothesis breadth $P_{50}$.}
\label{fig:baselines}
\end{figure*}

Table~\ref{tab:baselines} and Fig.~\ref{fig:baselines} report the head-to-head, under a \emph{semantic relay} protocol, where the sender perceives a scene, encodes a message with the codec under test, the message crosses the emulated channel, and the receiver must identify the sender's scene from the message alone. The JSON baseline remains more efficient in latency and payload size, requiring only a 120-byte message. The latent representation also leads to a sharper decoded belief (\(P_{50}=1.58\)) compared to the symbolic baselines (\(\approx2.25\)), while D3 confirms the preservation of conflicting hypotheses in \(92.75\%\) of trials. To summarize, AeroLat improves $\mathrm{SR}$ by $0.24$ over the best symbolic baseline, at the expense of a significantly larger transmitted payload.

\subsection{Ablations}\label{sec:ablations}
\begin{itemize}
    \item $\lambda$ sweep and the two fusion regimes
    Under evidence-refreshed dynamics (Fig.~\ref{fig:lambda}), $\lambda$ traces a clean trade-off: same-scene consensus rises $0.50\!\to\!0.84\!\to\!1.00$ as $\lambda$ goes $0\!\to\!0.4\!\to\!1.0$, while fused-state probe accuracy holds at $1.000$ up to $\lambda{=}0.3$, is $0.978$ at $\lambda{=}0.4$, and degrades beyond ($0.933$ at $0.5$; $0.489$ at $1.0$). Under pure \emph{gossip} dynamics (iterating fusion on already-fused states), the swarm converges to cross-scene consensus $1.0$ for \emph{any} $\lambda\ge0.1$ with fused accuracy collapsing to $0.489$.

    \item Component ablations:
        \begin{itemize}
            \item \emph{Whitening}: After experiments, it was found that probe accuracy remained unchanged ($0.915\to0.916$) while the mean off-diagonal cosine dropped from $0.999\to0.188$ on the full corpus geometry, not content, matching Table~\ref{tab:beforeafter}. 
            \item \emph{CLIP anchor}: With visual evidence removed from the prompt, probe accuracy falls to $0.18\approx$ chance versus $0.78$ with it. The results identified a collapse mechanism and demonstrate that scene-specific visual content originates from the visual anchor. The effect of the number of extracted tokens is shown in Fig.~\ref{fig:ksweep}, and the ablations in relation to the compressor loss are summarized in Table~\ref{tab:comp}.
        \end{itemize}
\end{itemize}

\begin{figure}[!t]
\centering
\includegraphics[width=\columnwidth]{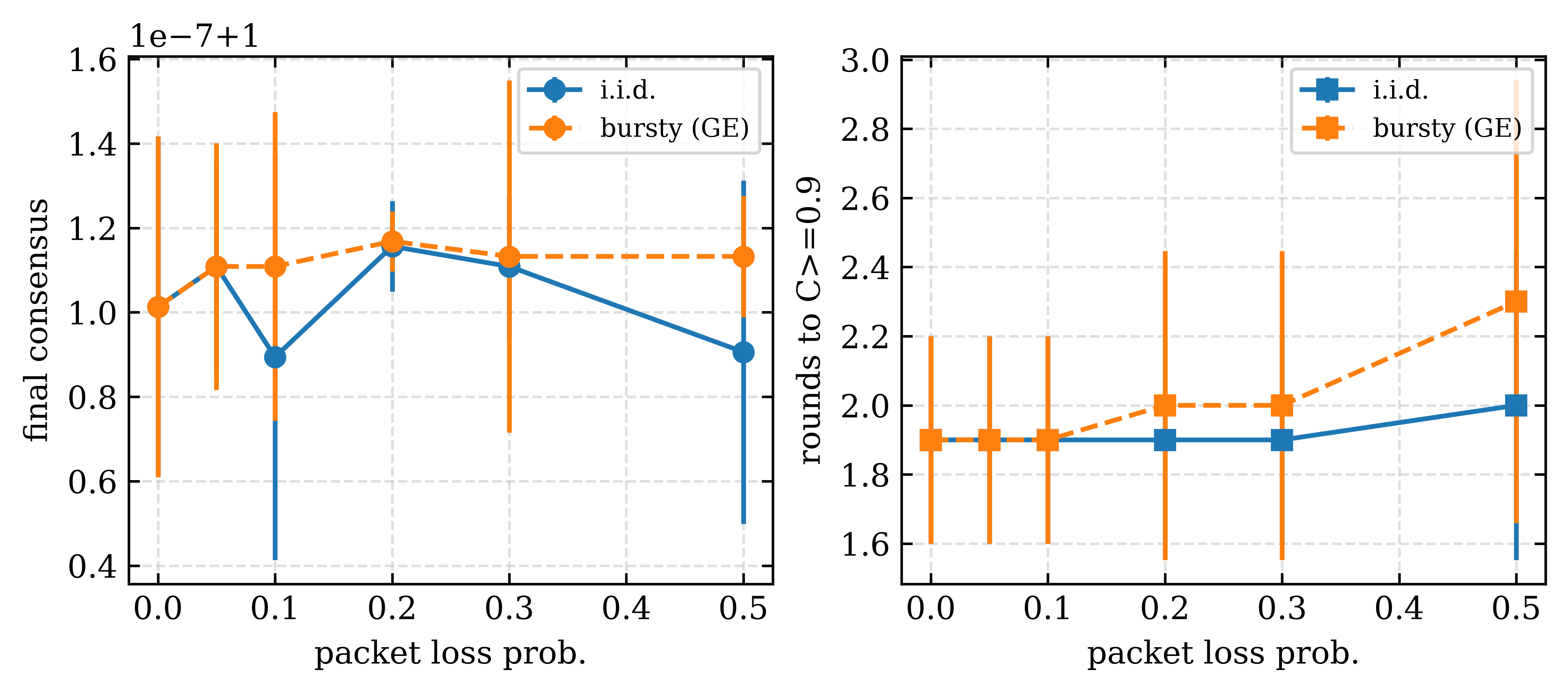}
\caption{Consensus under packet loss (i.i.d.\ and Gilbert--Elliott burst; $N{=}5$ fusion mesh, ten seeds). Final same-scene consensus is $1.00$ at every loss rate up to $50\%$; loss costs convergence \emph{time} only ($1.9\to2.3$ rounds).}
\label{fig:loss}
\end{figure}

\begin{figure}[!t]
\centering
\includegraphics[width=0.7\columnwidth]{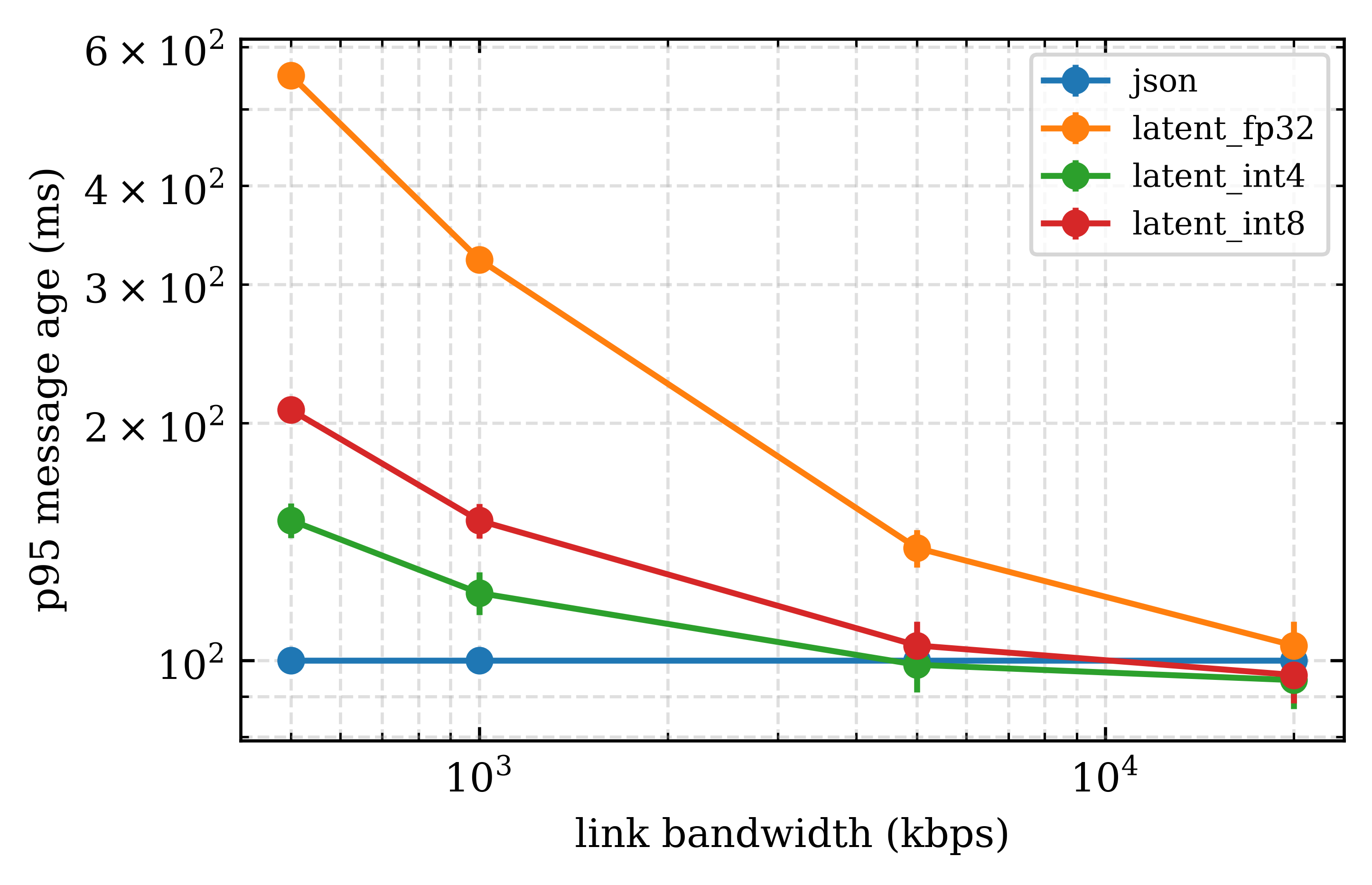}
\caption{Age of information under bandwidth caps ($50$\,ms link latency): p95 message age versus link rate per codec. At $500$\,kb/s, quantization moves the latent message from $552$\,ms (fp32) to $208$\,ms (int8) and $151$\,ms (int4), against JSON's $100$\,ms floor; at ${\ge}5$\,Mb/s all codecs converge.}
\label{fig:aoi}
\end{figure}

\begin{figure}[!t]
\centering
\includegraphics[width=\columnwidth]{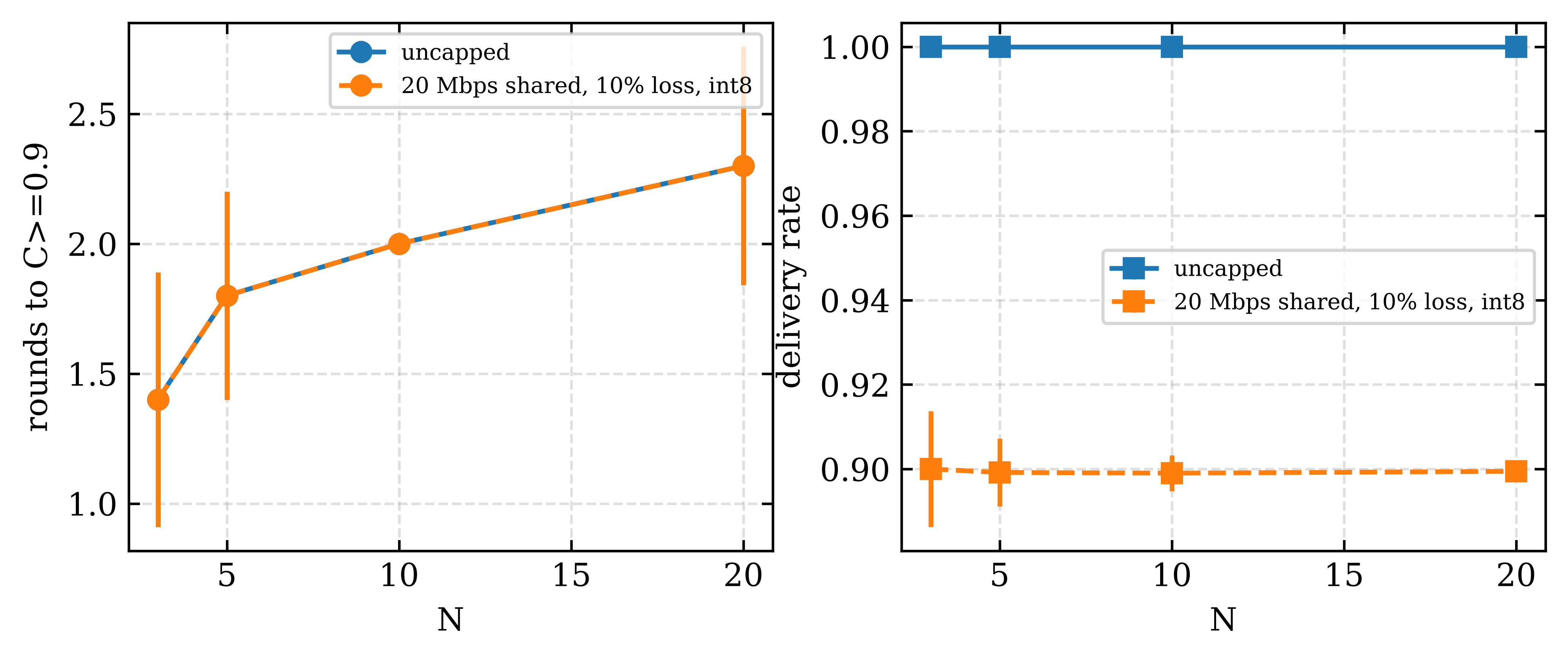}
\caption{Mesh scaling (ten seeds): rounds to $C\ge0.9$ grow sub-linearly ($1.4\to2.3$ for $N{=}3\to20$), and under a shared $20$\,Mb/s medium with int8 payloads and $10\%$ loss the mesh holds $0.90$ delivery with unchanged convergence at every $N$.}
\label{fig:scaling}
\end{figure}

\subsection{Mobility: Delay and Delivery Under Node Motion}\label{sec:mobility}

The channel, as depicted in Sec .~\ref {sec:channelmodel}, is static, but the aerial nodes are not. Henceforth, extend the link model with a standard block-fading treatment of mobility and sweep it offline. The results found show that int4 quantization is effectively \emph{task-lossless}. The mobility results present that the same compression is also crucial for reliable latent communication in time-varying channels. At a $0.9$ delivery floor, the fleet supports $N{=}4$ at fp32 and $N{=}10$ at int4---the $8\times$ rate reduction that costs nothing in task information buys a $2.5\times$ larger swarm. The result provides an operational interpretation of the offered-load trend directly in terms of the scalability and reliability constraints relevant to network design.

\subsection{Full-Stack SITL Campaign}\label{sec:sitl}
\begin{table*}[!t]
\caption{Eighteen-mission SITL campaign. One mission hang was reclaimed by the per-mission watchdog; $17/18$ completed nominally. Bold marks the values discussed in the text: the latent codec's success rate and the two failure signatures (CoT's mesh flooding, the no-whitening control's collapsed consensus).}
\label{tab:sitl}
\centering
\footnotesize
\begin{tabular}{|l|c|c|c|c|c|c|}
\hline
Configuration & Missions & SR & Encode ms (med/p95) & Whitened $C(t)$ & Delivery & Mesh delay \\
\hline
Symbolic CoT & 2 & $0.00$ & $3{,}137\,/\,3{,}464$ & $0.638\pm0.083$ & $1.00$ & $\mathbf{22.6}$\,\textbf{s} \\
JSON dispatch & 3 & $0.33\pm0.47$ & $71\,/\,79$ & $0.886\pm0.041$ & $1.00$ & $39$\,ms \\
Latent, dead drone $\{1,2,3\}$ & 3 & $0.67$ ($2/3$) & $114$--$120$ & $0.83\pm0.06$ & $1.00$ & $13$--$60$\,ms \\
Latent, $20\%$ loss + int8 + $50/25$\,ms jitter & 2 & $0.50$ & $117\,/\,128$ & $0.656\pm0.100$ & $0.80$ & $136$\,ms \\
Latent, \emph{no whitening} (control) & 2 & $0.00$ & $146\,/\,159$ & $\mathbf{0.999\pm0.000}$ & $1.00$ & $134$\,ms \\
\textbf{Latent} & 5 & $\mathbf{1.00}$ & $117\,/\,127$ & $0.678\pm0.094$ & $1.00$ & $51$\,ms \\
\hline
\end{tabular}
\end{table*}

\begin{figure}[!t]
\centering
\includegraphics[width=0.5\textwidth]{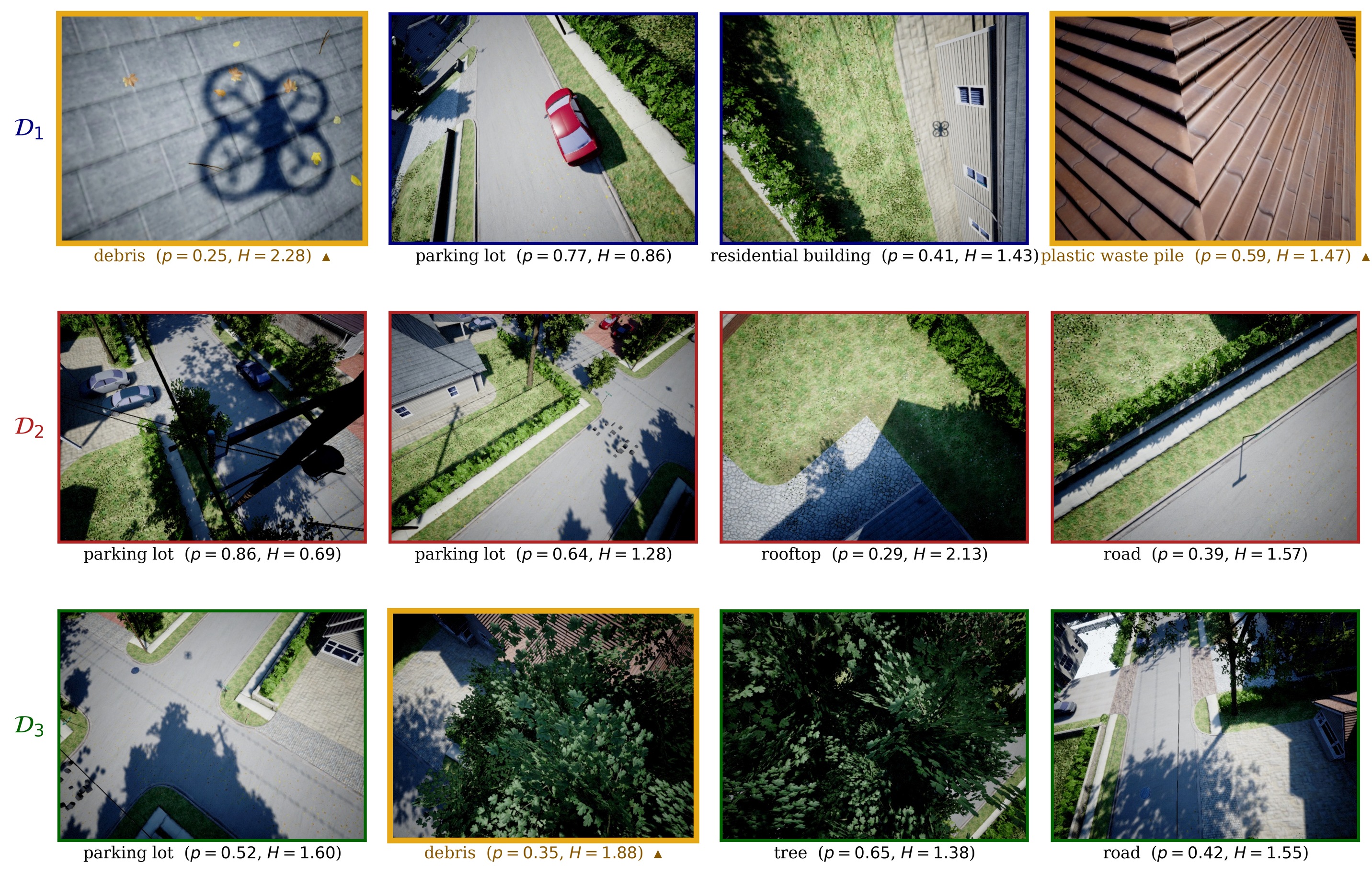}
\caption{The frame of what the drones actually see. The heterogeneity is the point: at any instant the swarm holds genuinely different, individually ambiguous views---$\mathcal{D}_1$ reads the spawned cluster as \emph{debris} at only $p{=}0.25$ with $H{=}2.28$ nats---and it is exactly this residual uncertainty that the whitened latent channel transports and the argmax-based JSON codec destroys .}
\label{fig:qualitative}
\end{figure}

Table~\ref{tab:sitl} reports the embodied campaign, and Fig.~\ref{fig:qualitative} shows representative onboard percepts from a capture mission. The findings from the Table~\ref{tab:sitl} and Fig.~\ref{fig:qualitative} are as follows:
\begin{enumerate}
\item \textit{The task result carries to flight.} AeroLat detected the anomaly in all five missions, compared to the JSON baseline, which detected the anomaly only in one of three runs, and CoT, which failed in both evaluated missions. The difference in performance is due to communication and inference overhead rather than to differences in visual sensing, even though all the stated methodologies use the same perception stack. In particular, CoT took about $3.1$ s to generate each message at the edge platform, increasing the mean message age to $22.6$ s compared with $51$ ms for AeroLat.

\item \textit{The collapse is reproduced live.} Thus, the whitened AeroLat representation was much more sensitive to changes in the visual environment.

\end{enumerate}

\subsection{Embodied Scaling and the Edge-Compute Ceiling}\label{sec:sitlscale}
\begin{table}[!t]
\caption{Embodied scaling campaign. The campaign is incremental by design: each size must succeed before the next is attempted, so the failure boundary is itself a measurement.}
\label{tab:scale}
\centering
\footnotesize
\setlength{\tabcolsep}{3.4pt}
\begin{tabular}{ccccccc}
\toprule
$N$ & Missions & SR & Enc.\ ms & Whitened $C(t)$ & $\erank$ & Msg age \\
\midrule
3  & 5 & $1.00$ & $117$ & $0.678\pm0.094$ & $1.9$ & $51$\,ms \\
5  & 3 & $1.00$ & $119$ & $0.649\pm0.070$ & $3.8$ & $78$\,ms \\
10 & 2 & $1.00$ & $148$ & $0.632\pm0.068$ & $7.6$ & $2.9$\,s \\
15 & $0/2$ & \multicolumn{5}{c}{watchdog $900$\,s / boot failure; $11/15$ airborne, $132$ frames} \\
\bottomrule
\end{tabular}
\end{table}

The emulated sweep establishes the communication load that the network can support. A separate flight campaign is used to measure the scalability of the full embodied stack. We attempted $N\in\{5,10,15,20\}$ incrementally (two seeded missions per size, frames, and per-tick latent dumps enabled, automatic abort once a size fails twice), on the same single-edge host that runs the simulator and all $N$ autopilots, and the shared perception-LLM pipeline simultaneously. There are four observations: 
\begin{enumerate}
\item \textit{The task metric is flat where flight completes.} Every completed mission at every size confirmed the anomaly ($\mathrm{SR}{=}1.00$), and mesh delivery remained $1.00$ throughout, across the entire campaign to $N{=}10$, not one latent packet was dropped.

\item \textit{No collapse at scale.} The effective rank of the circulating whitened states grows almost linearly with fleet size from \(1.9\) to \(3.8\) and \(7.6\), corresponding to approximately 0.75$N$. This indicates that adding drones adds \emph{dimensions} to the swarm's shared thought-space rather than redundant copies of a single thought, and that same-scene consensus remains within the discriminative band ($0.63$--$0.68$) rather than drifting toward the $0.999$ artifact.

\item \textit{The cost that grows is compute, not bandwidth.} Median encode latency rises from $117$ to $148$\,ms and, decisively, mean received-message age jumps from $78$\,ms at $N{=}5$ to $2.9$\,s at $N{=}10$ \emph{while delivery stays perfect}. The delay is inference queuing on the shared GPU, not the network. The anytime, staleness-aware fusion absorbs this; $SR$ is unharmed, but the trend identifies the true scaling bottleneck on edge silicon. It is the mirror image of the emulated result that the \emph{mesh} holds at $N{=}20$ as in Fig.~\ref{fig:scaling}. The embodiment shows that the \emph{processor} saturates first.
\end{enumerate}

\subsection{Comparison with the existing methodologies}\label{sec:sota}
Against Interlat, we share the extraction protocol and loss taxonomy but replace the function-call channel with an emulated lossy, noisy, band-limited network crossed by a live TCP mesh during flight, and we confront the collapse regime that embodied swarms. Against LatentMAS, we share the training-free ethos, but KV-cache working memory presupposes lossless co-located transport, whereas AeroLat's fixed $7{,}168$-dimensional state is engineered for a budgeted link with quantified distortion. Against the collapse literature~\cite{collapse}, we contribute the mechanism, a closed-form decentralized mitigation, and the observation that collapse splits into representational and dynamical channels. Against DeepSC-class semantic communication~\cite{deepsc,sana6g,kgsemcom}, we use the evaluation toolkit but transmit frozen LLM native states, achieving graceful degradation without training a codec. 
\section{Conclusions}\label{sec:conclusion}
AeroLat advances latent inter-agent communication from a software-benchmark mechanism to a network-realistic, statistically validated semantic communication system for UAV swarms. It identifies and mitigates representational collapse, explicitly models the communication impairments experienced by transmitted latent states, and quantifies the task-relevant information preserved across the channel.

 As part of future work, we will look at hardware-in-the-loop radio experiments to validate AeroLat under measured fading, interference, and contention. We also plan to extend the framework to heterogeneous UAV fleets via latent alignment among different frozen models.


\balance

\bibliographystyle{IEEEtran}
\bibliography{ref}

\end{document}